# Predictors of Social Distancing and Mask-Wearing Behavior: Panel Survey in Seven U.S. States[☆]

Plamen Nikolov[abcd]     Andreas Pape[a]     Ozlem Tonguc[a]     Charlotte Williams[a]

September 2020

**Abstract**

This paper presents preliminary summary results from a longitudinal study of participants in seven U.S. states during the COVID-19 pandemic. In addition to standard socio-economic characteristics, we collect data on various economic preference parameters: time, risk, and social preferences, and risk perception biases. We pay special attention to predictors that are both important drivers of social distancing and are potentially malleable and susceptible to policy levers. We note three important findings: (1) demographic characteristics exert the largest influence on social distancing measures and mask-wearing, (2) we show that individual risk perception and cognitive biases exert a critical role in influencing the decision to adopt social distancing measures, (3) we identify important demographic groups that are most susceptible to changing their social distancing behaviors. These findings can help inform the design of policy interventions regarding targeting specific demographic groups, which can help reduce the transmission speed of the COVID-19 virus. (*JEL* I11, I12, I18, D81, D91, D62, D64)

We thank Ariunzaya Oktyabri, Alan Adelman, Ancilla Inocencio, Xu Wang, Sharon Itin, Dylan Saez, and Hannah Werner for outstanding research support. We thank Jacob Hanna, Andrea Youngken, Jefferey Yan, Anthony DeMaria, Farris Bienstock, and Zachary Pinto for assisting with the data collection process. We thank Zoe McLaren, Lulu Garcia-Navarro, Hanna Tesfasyone, Matthew Bonci, Robin Schilling, Susan Wolcott, Nusrat Jimi, Agnitra Roy Choudhury, and Leila Salarpour for constructive feedback and helpful comments. This research was made possible thanks to grants from The State University of New York Research Seed Grant Program, Binghamton University's SSEL, and the Health Sciences and Sustainable Communities Transdisciplinary Areas of Excellence. We gratefully acknowledge their support. All errors are our own, please let us know about them.

[☆]Corresponding Author: Plamen Nikolov, Department of Economics, State University of New York (Binghamton), Department of Economics, 4400 Vestal Parkway East, Binghamton, NY 13902, USA. Email: pnikolov@post.harvard.edu

[a] State University of New York (at Binghamton)
[b] IZA Institute of Labor Economics
[c] Harvard Institute for Quantitative Social Science
[d] Global Labor Organization

**I. Introduction**

Absent a vaccine or antiretroviral drugs to curb the spread of the COVID-19 virus, the only immediately available and possibly effective tools to decrease transmission are preventive measures, via individual behaviors. Government guidelines quickly pointed to two specific measures: social distancing and mask-wearing.[1,2] In this paper, we examine what factors predict the tendency to adopt social distancing measures. Although many Americans have adopted these measures, many have not opted to do so (Saad, 2020). For social distancing to be effective against the spread of the disease, its adoption must be widespread and consistent.

What are the most important drivers for social distancing? Are any of them subject to policy changes? We use longitudinal data, based on an online field experiment from seven U.S. states collected between May and July 2020, to address these questions. We recruited participants via online platforms who completed weekly surveys. Participants completed five surveys (two in the first week: an initial baseline and a follow-up survey, and surveys for each consecutive week for three weeks). The weekly surveys collected data on social distancing behaviors, general health, social preferences, time preferences, risk attitudes, cognitive assessment, attitudes towards COVID-19, perception biases, and personality traits. Based on four individual-level proxies of social distancing –days outside of one's home in the past week, whether one left one's home, whether one maintained at least six feet of distance, and the number non-essential reasons for leaving one's home – we create a comprehensive social distancing index. Using the survey data, we take a reduced-form approach as we attempt to shed light on the relationship between the above factors and the demand for social distancing and mask-wearing.

The classical economic model of individual demand for health, based on Grossman (1972), posits that individuals view health and preventive health measures as investments.

---

[1] On March 16th, 2020, President Donald Trump issued the first White House Coronavirus guidelines in which it was officially recommended that citizens avoid gatherings in groups of more than ten people at a national level (White House, 2020).
[2] Social distancing describes the practice of keeping a safe distance of at least six feet between yourself and those not from your household (CDCb, 2020). This guideline aims to reduce the potential for physical transmission of the virus from the infected to the non-infected. This practice is recommended in conjunction with the use of personal protective equipment (PPE) (face masks, gloves, etc.). PPE aims to limit respiratory transmission of the disease through breathing, coughing, and talking.

Consumers adopt better health behaviors by comparing the benefits and costs associated with these behaviors. This framework places the role of individual incentives front and center: viewing health in an inter-temporal framework, individuals make health investment decisions (e.g., exercise, nutrition) in the present to maximize their lifetime well-being. Changes to individual behavior in the present are associated with disutility (e.g., most changes to habits and behavior require some or a considerable degree of effort), and the adoption of new health behaviors requires a considerable amount of time investment or additional monetary costs.[3] Social distancing and mask-wearing are no different. Understanding how individuals make decisions regarding social distancing is further complicated by the fact that COVID-19 is an infectious disease and that preventive measures affect not only individual health, but also the health of others. In other words, individual decisions regarding optimal preventive measures could be socially suboptimal if people only care about their own health and not about how their behaviors affect the spread of disease to others, a classic case of a negative externality. Using data on various economic preference parameters, such as willingness to take on gambles, willingness to delay gratification, pro-social attitudes, and personality traits, we shed light on the determinants of individual decisions to adopt social distancing or mask-wearing. We pay special attention to factors that either exert a considerable influence on social distancing or are potentially malleable and susceptible to policy levers.

In addition to individual economic preferences, we focus on the impact of two perception-related variables. The perceptual biases that we examine, risk misperception, and exponential growth bias (EGB), can affect decision-making and individual behaviors.[4] By risk misperception, we refer to the difference between the subjective assessment of a disease risk versus the objective disease severity risk. In the context of COVID-19 risks, we specifically focus on the role of the mortality risk (or the so-called case fatality rate). We also examine the effect of EGB, the psychological phenomenon in which individuals tend to linearize exponential growth. This tendency leads to underestimating the future growth of a particular variable, which leads to suboptimal decisions. In the context of COVID-19, exhibiting EGB simply means that people underestimate the speed at which the epidemic spreads. Although the implications of EGB for individuals making health decisions are immense, previous studies have exclusively focused on the welfare consequences of the phenomenon in the context of savings and debt (Wagenaar and Sagaria 1975; Wagenaar and Timmers 1979; Keren 1983; Benzion, Granot and Yagil 1992; Eisenstein and Hoch 2007; Alemberg and Gerdes 2012). In this paper, we examine whether people exhibit exponential growth bias in the health domain and whether its presence relates to lower willingness to adopt social distancing measures.[5]

---

[3] Strulik and Trimborn (2018) find that individuals continuously revise their life behaviors in order to increase health investment as they are driven by the desire for a long life, they note that these revisions are failed to be adhered to due to changing external biases.

[4] Conceptually, at least two behavioral biases predict that people may not engage in optimal levels in preventive health measures, such as social distancing: present bias (PB) and exponential growth bias (EGB). Present bias (PB) is the tendency to underweight future utility relative to present utility in a dynamically inconsistent way (Strotz 1955; Laibson 1997; O'Donoghue and Rabin 1999).[4] Previous studies document that in intertemporal models, present bias can lead to lower demand for health relative to an unbiased person who shares the same long-run discount factor (Dupas 2011, Laibson 1997, Angeletos et al. 2001). A person with present-biased preferences may intend to invest more in health in the future but never do so (Dupas 2011).

Understanding the mechanisms behind what drives suboptimal health decisions and social distancing, in particular, can inform the design of better policies. Although many factors may affect social distancing, we explore the possibility that for at least some people, the low adoption may be the result of low demand due to low perceived risks associated with COVID-19. If perceptual biases are the main reason behind the low adoption of social distancing, reducing such biases could lead to better health behaviors and higher social welfare. It is especially important to understand the role policies can play in mitigating such potential biases. Identifying which biases influence the decision to social distance can help inform the design of better policy interventions, from a welfare standpoint.

Our analyses yield several interesting results. First, and controlling for other factors, political affiliation is a strong predictor of individual views on how various governmental levels handled the response to the epidemic. Democrats and Independents are more likely to view Federal, State, and Local government responses as not sufficiently aggressive compared to the views on the same topic by Republicans. This finding is consistent with recent survey evidence from smartphone data (Allcott et al., 2020), documenting that areas with more Republicans engaged in less social distancing. From a theoretical standpoint, the fact that political affiliation, in contrast to economic preference parameters, is a strong driver of the decision to social distance could demonstrate the usefulness of economic models related to the formation of social norms in the context of health behaviors.[6]

---

[5] Levy and Tasoff (2016) show in the context of savings behaviors that EGB can lead to under-saving for retirement under certain conditions. A large body of evidence suggests that this bias is widespread and robust (Wagenaar and Sagaria 1975; Wagenaar and Timmers 1979; Keren 1983; Benzion, Granot and Yagil 1992; Eisenstein and Hoch 2007; Almenberg and Gerdes 2012).

Second, of all the variables included in our analyses, demographic characteristics are the greatest drivers of social distancing measures. Information on demographic factors that predict social distancing and mask-wearing can inform targeting of potential policies. Specifically, being Caucasian, a Republican affiliation, and higher household income are associated with lower adoption of social distancing measures. Living in a densely populated area, being currently married, and being highly concerned about own health are positively associated with higher rates of social distancing. Demographic characteristics also influence mask-wearing. Higher age and being female are both associated with more mask-wearing. In contrast, being Caucasian is associated with lower rates of mask-wearing. Among the economic preference factors, we find that exhibiting higher risk aversion is a strong predictor of social distancing and mask-wearing. We find no evidence of time preference per se influencing social distancing measures. However, we find a surprising relationship between time preferences and mask-wearing: individuals who exhibit higher patience (higher willingness to delay gratification) are less likely to wear a mask.

Third, in addition to demographic and economic factors, we specifically examine the roles that perception and cognitive biases play in influencing the decision to adopt social distancing measures. We estimate the prevalence of the exponential growth bias in our sample. Numerous studies have addressed the theoretical relationship between standard time preferences and health behaviors (Grossman, 2017). However, surprisingly few attempts focus on the empirical relationship between EGB and health behaviors. Based on our sample, we establish an empirical link between the tendency to exhibit exponential growth bias and social distancing. We control for income, education, risk preference, time preference, a measure of cognitive skills, and a host of other characteristics, and find that that EGB is a statistically significant predictor of social distancing. On the aggregate social distancing index, people who misunderstand exponential growth (i.e., exhibit higher exponential growth bias) report a lower tendency to social distance. We also find suggestive evidence that individuals who misperceive the risk associated with COVID-19 are less likely to wear a mask.

---

[6] Social norms provide informal rules that govern our actions within groups across various situations. A key feature of a social norm is the desire to conform to the majority in a group, as noted by Bicchieri (2017, p. 35). Brock and Durlauf (2001) model the development of norms within a group; building on the model of Brock and Durlauf (2001), Smerdon et al. (2020) design an experiment in which they investigate the role that information about others' preferences on the development of "bad" norms. They find that the strength of social interactions is a critical factor for a bad norm to persist.

We note several important limitations of our analyses. The findings reported in our study rely on econometric specifications that should be interpreted with some caution from a causal standpoint. Although we find important predictors of social distancing behaviors, the variation in these predictors is not driven by an exogenous source of variation. However, many of the predictors are either demographic characteristics or economic preference characteristics that recent empirical studies have documented to exhibit stability over time: for risk preferences (Dohmen, Lehmann, and Pignatti, 2016; Chuang and Schechter, 2015; Nikolov, 2018), time preferences (Meier and Sprenger, 2015), personality traits (Roberts, 2009), or social preferences (Carlsson et al., 2014). Furthermore, establishing a causal relationship between economic preference parameters and social distancing measures would be a challenging task because behavioral parameters cannot be exogenously varied. Finally, our analysis relies on a small sample from seven states that exhibits important differences compared to the general population. Therefore, there may be difficulty extrapolating our findings to the general populations or other settings.

The structure of the paper is as follows. Section II describes the survey data and specific survey measures. Section III presents analyses on the most important predictors of adherence to social distancing and mask-wearing. Section IV discusses key takeaways from the analysis and provides some concluding remarks.

## II. Design and Survey Data

### A. Study Design and Survey Data

To examine the determinants of social distancing behavior, we conducted a longitudinal survey as part of an ongoing online field experiment (results from which we do not report here). The field experiment aimed to examine how random assignment to informational treatment influenced economic preference parameters and subsequent tendency to adopt social distancing and mask-wearing behaviors. Requirements to participate in the experimental study were: 1) being 18 years of age or over, and 2) current residence in one of seven states[7]: Connecticut, Florida, Maine, Maryland, Massachusetts, New Jersey, and New York, and 3) consent to be contacted by the study team for follow-up purposes.[8]

Recruitment and data collection for the study was executed remotely. Starting in May 2020, participants were invited to participate in the study through recruitment ads posted on the Binghamton University Experimental Economics Lab, online postings, and word of mouth. Upon invitation to participate in the study, individuals filled out a brief online screening form to ensure they met the study's eligibility criteria. Once enrolled, participants were invited to complete weekly surveys for seven weeks. The surveys collected information regarding social distancing behaviors, general health, social, time, and risk preferences, and exponential growth knowledge at that particular period in time.[9] Participants completed five surveys (two in the first week: an initial baseline and a follow-up survey, and three weekly surveys for three consecutive weeks). The follow-up surveys covered questions, also elicited during the baseline, to enable an analysis of changes in behaviors over the study period.

The baseline study instrument comprises the following modules: (a) socio-demographic information, (b) mask-wearing and social distancing, (c) exponential growth assessment, (d) social, time, and risk preferences, (e) health outcomes, (f) cognitive assessment, and (g) attitudes. We detail the definition of the critical variables in Section II.C.

We first report, in Table 1, the average characteristics of the study participants. Our overall sample is 42.40 years of age. The study sample is majority female, a split of 713 females and 188 males. The male sample reports a lower average age as compared to the female sample, with a reported average age of 32.99 and 44.88 years, respectively. The sample is primarily of Caucasian ethnicity, with an annual household income of approximately $116,000. In terms of educational attainment, over half of the participants (61 percent) report receiving a degree from a four-year college program. Approximately a tenth (nine percent) of the sample is obtaining an educational degree. In terms of employment, 48 percent of the sample reports being currently employed. It is worth noting that in terms of risk

---

[7] These states collected daily data updates on the number of cases and number of deaths related to COVID-19 within state borders at the county level.
[8] Weekly follow-up phone interviews were conducted with respondents where the research team discovered inconsistency in responses.
[9] Data collection was executed on the oTree platform, designed by Chen et al. (2016). Participants received incentive payments related to the incentive-based measurements of specific survey tasks (e.g., risk preferences and exponential growth bias measurements).

and time preferences, our sample was on average risk-neutral and consisted of individuals who are more likely to be patient than impatient regarding financial decisions.[10]

[Insert Table 1]

Table 2 reports the summary characteristics by the state of residence. The majority of the sample resides in New York State. The average participant that resides in New York is significantly younger and more likely to be a full-time student than those in the other six states. Although the majority of the sample is Caucasian, it is interesting to point out that the subsample residing in New York is the most ethnically diverse, with 4 percent of participants reporting to be African American. In contrast, Florida and Pennsylvania are the least diverse states in the sample, with no reported African Americans. We also report political affiliation by the state of residence: Florida (0.18) and Pennsylvania (0.18) tend to have the most Republican-voting participants in our sample. In contrast, participants in the following states of residence are more likely to report Democratic affiliation: Connecticut (.69), Massachusetts (.67), Maryland (.67), and New Jersey (.68).

[Insert Table 2]

### B. Social Distancing and Mask Wearing Behaviors

Since the primary objective of the study is understanding the determinants of social distancing and mask-wearing behaviors, we collect the various proxies of social distancing collected.

***Proxies of Social Distancing***. We collect data on social distancing and mask-wearing behaviors. We rely on four different proxy measures of social distancing: (1) a quantitative measure of the days outside of one's home in the past week, (2) whether one left one's home, (3) whether one maintained at least six feet of distance, and (4) a quantitative measure of non-essential reasons for leaving one's home. We measure the number of days one indicated having been outside of one's home based on the following question: "during the past seven days, how many days did you leave home?" We quantify this variable using a 7-point scale; 7 indicates that a person left the house every day in the past seven days. We measure whether one left one's home the day prior to taking the survey with a binary indicator, set to 1 if the person indicated they left the house.[11] We measure whether one maintained two meters (or six feet) from other people also based on a binary variable, set to 1 if the person did not maintain at least six feet from other people. Our final proxy measures the number of non-essential outings outside of the house.

---

[10] The risk aversion index, based on the switching point, is measured from 1 to 10 (1 indicates a very risk averse respondent; 10 a very risk-tolerant respondent).

Based on these four variables, we created an aggregated social distancing index which was generated based on a principle component procedure. The social distancing index is continuous and negatively coded (i.e., an increase in the index denotes less social distancing). Although we report information on how factors correlate with the primary index, we also report analyses based on the distinct proxy indicators of social distancing discussed above.

We report, in Table 3, how these proxies of social distancing (measured at the baseline[12]) vary by socio-economic characteristics.

[Insert Table 3]

We detect some differences by race for two specific proxies of social distancing. African Americans report a higher number of non-essential outings per week than Caucasians, although Caucasians leave the house significantly more days per week than African Americans, 3.22 and 0.21 days, respectively. Despite the difference in the specific measures, the overall social distancing index is very similar for both ethnicities at 0.07 for Caucasians and 0.08 for African Americans.

---

[11] For the variable "did not stay home," we used the following survey question: "In the past week, I stayed at home". Survey respondents responded on a scale from 0 to 100 (in increments of 10) with 0 indicating "does not apply at all" and 100 indicating "applies very much," with values of 100 indicating complete adherence (=1 for the binary indicator) on this proxy measure. For the variable "did not maintain six feet," we used the following question: "In the past week, I kept a distance of at least two meters (six feet) to other people." Values of 100 (out of a scale from 0 to 100) indicated completed adherence to the six-feet rule and were set to 1 for the binary indicator for this proxy measure. The number of non-essential outings was based on survey responses for the following question: "What are reasons for you to leave your home (select all that apply)?" Respondents could select among 13 choices regarding reasons to leave the home spanning from going to work and exercising to getting bored and exercising their freedom. We classified the following responses as "non-essential" reasons: getting tired of being inside the home, getting bored, getting adrenaline (from breaking the law), and exercising my freedom. The continuous measure can take on values up to 4.

[12] The baseline survey was collected prior to the Black Lives Matter protests, which started the last week of May 2020.

***Proxies of Mask-Wearing***. We also collected information on mask-wearing. We measure mask-wearing behaviors with two binary indicators. The first indicator measures whether a person wears a mask (or personal protective equipment) at all times outside of the home or not. The second indicator measures whether a person reported wearing a mask or personal protective equipment (PPE) when conducting daily business, shopping, or running errands.[13] Although the two indicators are related, the second indicator captures behaviors when a person wears a mask only when conducting essential activities and not necessarily wearing it at all times.

Table 4 reports descriptive analysis on mask-wearing behaviors by demographic characteristics: Caucasians tend to wear masks more frequently than African Americans in our sample when accounting for both daily use and use when shopping. However, over the two weeks (between the baseline and the second follow-up wave), African Americans indicated a considerably higher increase in mask usage than the change in mask-wearing behaviors among Caucasians.[14]

[Insert Table 4]

The results reported in Tables 3 and 4 show that Republicans in our sample tend to adopt social distancing significantly less than Democrats or Independents. Individuals who self-identify as Republican, and relative to individuals who self-identify with other political parties, report lower mask-wearing (both on the daily measure and when shopping). These results are consistent with a recent report based on observational data from Allcott et al. (2020). Average social distancing (based on the comprehensive index) increases with education level. Individuals with a high school diploma only social distance less, whereas individuals with a graduate school degree report social distancing at higher rates. Notably, this pattern among education groups does not align with the adoption of mask-wearing; graduate degree holders tend to wear masks the least.[15] We return to the issue of other correlates of social distancing and mask-wearing, once we account for other individual-level factors, such as individual economic preferences.

---

[13] The two mask-wearing binary indicators were based on two survey questions. The first question was based on a survey question "When I leave my home, I use personal protective equipment." Individuals indicating complete adherence (responses =100) were coded as wearing a mask. The second question was based on the following question "Under which circumstances do you use any personal protective equipment (PPE) (such as face mask and gloves)?" When the respondent selected the option "Daily business for example shopping, mailing, pharmacy", the indicator for mask-wearing when shopping was set to 1, and 0 otherwise. Data for this indicator was only available during the baseline week as participants were not asked the same question in the follow-up surveys. As a result, we are unable to subsequent behavior changes for this question only.

[14] Our sample based on changes in the variables is smaller due to attrition of subjects.

### C. Survey Data of Economic Preferences

We also collect rich information on an array of economic preferences variables, such as risk preferences, time preferences, social preferences, hyperbolic time-inconsistent behavior, and the tendency to exhibit exponential growth bias.

***Time Preferences.*** Given the conceptual link between economic models on health investments and time preferences, we elicit respondents' time preferences. As is common in the related literature, we measure time preferences by asking participants to choose either a smaller amount of monetary compensation sooner or a larger amount in the future (Tversky and Kahneman 1986; Benzion et al., 1989; Shelley 1993, Ashraf et al., 2006).[16]

We use two distinct *staircase* procedures.[17] The procedure presented the same questions for two assessments, with one presented in the near frame and one in the distant temporal frame. For the near-frame question, we elicit the respondent's willingness to choose the smaller reward immediately versus receiving a larger reward in 12 months. Participants were asked to choose either $160 today or $246 in twelve months. If they chose the smaller amount today, the next question would offer them the same amount today ($160), but an amount higher than $246 in 12 months. If they chose $246 in 12 months, the next question offers them the same amount today ($160), and an amount smaller than $246 in 12 months. The participant responds to a maximum of five questions, and each time their future payoff is adjusted based on earlier responses. The respondent is then asked the same initial question regarding a future time frame (12 months versus 24 months but with the same monetary rewards) to identify time-preference reversals. Based on the multiple time-preference questions, we created a continuous index from 0 to 32, with increasing numbers indicating

---

[15] A recent Gallup poll contradicts this pattern and finds that college graduates tend to wear masks more compared to individuals without a college degree (Brenan, 2020).
[16] Anderson et al. (2002) and Frederik et al. (2002) review elicitation methods for time preferences.
[17] The procedures used was based on Falk et al. (2016) and Holzmeister (2017). The staircase procedure streamlines the elicitation process by requiring considerably fewer questions, in contrast to MPL elicitations.

higher patience (32 on the index denoted extreme tendency to delay gratification and preference for the delayed reward). If an individual chooses to take a smaller amount of money today for the first question, we define them as impatient in the near frame (with the same process repeated in the distant frame). To do so, we use binary indicators based on near-frame questions. Specifically, we denote individuals as "very patient" (set to 1) if they have a score of 32 on the preference index. We use a binary indicator to denote individuals as "very impatient" (the binary indicator is set to 1) if the individual has a score of 1 on the preference index).

Based on the time-preference indices of the near and distant frames, we can also observe time-inconsistent choices. We define a participant to be "hyperbolic time-inconsistent" based on the choice of the immediate reward in the near-term frame combined with the choice of the delayed reward in the distant frame since the implied discount rate in the near-term frame is higher than that of the distant frame. We also identify inconsistencies in the other direction, where individuals are patient *now* but in 12 months are *not* willing to wait; we refer to these as individuals as "patient now and impatient later."[18]

***Risk Preferences.*** We elicit risk tolerance based on an incentivized risk elicitation task, the multiple price list (MPL), based on Holt and Laury (2002)[19]. In such activity, participants are prompted with two options, Option A and Option B, with Option B being "riskier."[20] The individual chooses which option they prefer. When the probability associated with the high payoffs increases sufficiently (as the individual moves down the table), a risk-averse person should cross over to Option B. We use this switching point to Option B to quantify one's risk tolerance—participants with a higher risk-tolerance switch at lower switching row values. In contrast, more risk-averse participants tend to switch at higher switching row values.[21]

---

[18] Individuals are defined as time-inconsistent if their switching row value is lower in the near frame than in the far frame (indicating they are more patient in the near frame than in the far frame).
[19] Charness et al. (2013) reviews limitations of non-incentivized tasks for risk-preference elicitations.
[20] In the MPL, participants are presented with two lists of options: Option A and Option B. Each option represents a gamble: Option A offers 1/10 a chance of winning $2 and 9/10 chance of winning $1.60 while Option B offers 1/10 chance of winning $3.85 and 9/10 chance of winning $0.10. The participants marked which of the two options they would prefer, with Option B indicating an individual with a higher tolerance for risk. The questions continue with the chances of winning the higher dollar value in both rising until a 100% chance is offered for winning both $2 and $3.85 in Option A and B respectively. The question at which the player switches from Option A to Option B is denoted as their switching row.

***Objective COVID-19 Mortality Risk and Perceived Mortality Risks.*** We elicited subjective assessments of mortality risks associated with COVID-19. Quantifying perceptions of risk related to COVID-19 is difficult, especially with young respondents. Valuable summaries of methods for eliciting expectations for a range of outcomes can be found in Manski (2004) and Delavande, Gine, and McKenzie (2008). Therefore, in this survey, we only elicited risk perceptions based on simple questions about the perceived risk of mortality related to COVID-19. Individuals were prompted to provide a subjective assessment regarding the fatality rate of COVID-19. Specifically, each respondent reported a subjective estimate for a hypothetical set of 1000 individuals in their county of residence who are infected with COVID-19. We use this proxy measure to serve as a subjective assessment of the so-called case fatality rate (CFR), measuring the severity of a disease. The CFR is the proportion of people who die from a specified disease among all individuals diagnosed with the disease over a certain period.

These simple questions eliciting the subjective assessment of the CFR rate have several evident and significant limitations. They are not precise in specifying the meaning of "expected." The measure does not elicit uncertainty. The questions also deal with an abstract and hypothetical situation.

Based on data obtained from the Centers for Disease Control and Prevention (CDC), we also collect data on the objective risk, capturing the weekly CFR rate at the county level. This data is merged with our longitudinal panel survey based on the county of residence of each respondent.

We also calculate the difference between the subjective perception and the objective risk, a measure which we call subjective misperception (positive values indicate over-perception of the CFR rate; negative values indicate under-perception of the CFR rate). In summary, the misperception measure indicates whether personal subjective assessment about the mortality risk considerably deviates from the objective risk data related to COVID-19.

---

[21] The elicitation method is complex for some participants as noted in Dave et al. (2010) and Holt and Laury (2002). Therefore, some participants switched more than once between the two options. We address this issue following the approach in Engel and Kirchkamp (2016).

***Social Preferences.*** We also collect data on two proxies of social preferences: positive reciprocity and altruism.[22] Positive reciprocity is the tendency for an individual who receives a positive benefit from an action to reciprocate with an action that has at least an equal amount of complimentary benefit (Suranovic, 2001).[23] We proxy positive reciprocity with two measures, a quantitative and a qualitative measure. The qualitative question asks individuals to self-reflect on how much the following statement applies to them: 'When someone does me a favor, I am willing to return it." The quantitative question poses a hypothetical scenario in which the participant is in an area they do not know and their phone is about to run out of battery. The participant is then prompted to indicate the monetary value of a gift they would be willing to gift to a hypothetical stranger who presented them with a charging cable worth $40.

We also collect data on proxies of altruism, the concept of one's capacity for unconditional kindness. Altruism does not reflect reciprocity in response to someone else's altruism (Fehr and Gächter, 2000). We proxy altruism with a qualitative survey question; respondents were asked to self-reflect on how much the following statement applies to them: "How willing are you to give to good causes without expecting anything in return?"[24]

***Exponential Growth Bias.*** Exponential growth bias (EGB) is the tendency of people to exhibit a perceptual bias to underestimate exponential growth. The phenomenon refers to peoples' tendency to linearize exponential growth. Individuals who exhibit EGB underestimate the future value of a variable growing at a constant rate. Previous economics studies explore the importance of this phenomenon for savings decisions and for underestimating exponential growth due to compound interest (i.e., a person who exhibits EGB underestimates the returns to savings and the costs of holding debt).[25] Because people who exhibit EGB underestimate the speed at which the spread of the epidemic grows, we collect data to proxy EGB so we can examine whether the magnitude of the EGB is related to the adoption of social distancing measures.

---

[22] Social preferences were elicited based on the procedures adopted by Falk et al. (2016).
[23] We also measure negative reciprocity. Positive and negative reciprocity are weakly correlated (Dohmen et al., 2008).
[24] We collect information based on the following question: "Today you unexpectedly received 1,600 U.S.dollars. How much of this amount would you donate to a good cause?"
[25] Starting with Stango and Zinman (2009), several studies find evidence of exponential growth bias (EGB) in financial decisions based on laboratory or field experiments. Levy and Tasoff (2015) document that EGB can lead to under-saving for retirement. Other studies document the existence of this bias in numerous settings (Wagenaar and Sagaria 1975; Wagenaar and Timmers 1979; Keren 1983; Benzion, Granot and Yagil 1992; Eisenstein and Hoch 2007; Almenberg and Gerdes 2012). No empirical evidence exists on the existence of EGB in the context of disease spread.

To proxy the prevalence of exponential growth bias, we ask participants a series of five questions similar to the approach adopted by Stango and Zinman (2009) and Levy and Tasoff (2016). Questions covered growth phenomena in several domains: epidemic growth, population growth, and financial asset growth. In the elicitation method, participants are asked to calculate the final value given the initial value, growth rate, and period.[26] For example, in the context of financial assets, we asked participants to calculate how much an investment of a certain amount would be worth after a specified period with a given interest. Similar questions are asked about the growth of a disease: information is provided about the initial number of people infected by an infectious disease and the number of new infections in a specified period (growth rate); participants are prompted to provide their assessment on the number of total infections within a given timeframe. Among the five questions, three questions referred to the growth of an amount in a bank account at a given interest rate, and two questions referred to the growth of a disease spread. Each response is marked as correct or not. Based on standardized measures from these five binary indicators, we generate a comprehensive index. We use the aggregated EGB index (higher and more positive values are associated with a higher tendency for EGB) to examine how it relates to the study outcomes.

***Assessment of Cognitive Skills and Personality Traits.*** We use an abbreviated version of the Raven's Progressive Matrices. The Raven test, which consists of non-verbal multiple choice questions, is recognized as a leading measure of analytic intelligence (Carpenter et al., 1990; Gray and Thompson, 2004). Each question asks the subjects to identify the missing element that completes a visual pattern. The test was not incentivized, a method consistent with the approach adopted in recent psychology studies (Gill and Prowse, 2014).[27]

We also collected data on the "Big Five" personality traits: openness, conscientiousness, extraversion, agreeableness, and neuroticism (John and Srivastava, 1999).[28] Cobb-Clark and Schurer (2012) show that personality traits can influence economic decisions.

---

[26] Subjects were not prohibited from using calculators or online tools to help them answer the questions.
[27] The Raven's cognitive assessment consisted of four tasks with time allocation 120 seconds. Based on the number of correct answers and the time finished, we generated a variable for the number of correct responses per time taken to complete the task.
[28] Openness measures creativity and imagination, conscientiousness measures care and self-discipline, extraversion measures sociability and energy, agreeableness measures tendency to adhere to social harmony, and neuroticism measures negativity and

## III. Factors Predicting Social Distancing Behaviors

In this Section, we examine the participants' views on the government's responses by political affiliation. For these analyses, we use data from the baseline. We also use the panel feature of the survey design to examine changes in individual behaviors. Specifically, we analyze predictors of the social distancing measures, with a particular focus on the economic preferences and the perception-related variables to predict the decision to adopt social distancing measures.

### A. Views on Government Response by Political Affiliation

We analyze the respondents' views, by political affiliation, of the government's response to COVID-19. Respondents provided their assessment on the reaction of various levels of the government based on the following question: "Do you think the reaction of the Federal/State/Local government to the current coronavirus outbreak is appropriate, too extreme, or not sufficient?" We measure the reaction to the government's response to COVID-19, on a scale of one to five, with 1 being "*much too extreme*" and five being "*not sufficient at all.*" Based on the scale used for this question, a response of 3 indicated "*the reaction is appropriate*" for all questions.

Table 5 reports the aggregated results. The most favorable views are associated with the State government's response. In contrast, the least favorable views relate to how the Federal government handled the response to the epidemic.

[Insert Table 5]

Political affiliation is a strong predictor of individual views on how various governmental levels handled the response to the epidemic. Democrats and Independents are

---

self-doubt (John and Srivastava, 1999). We elicited the traits based on a Likert-style scale using the following prompts: "I see myself as (trait)." Each trait is related to two survey questions, which then are standardized and used to generate a continuous index measurement for each trait.

more likely to view Federal, State, and Local government responses and the general public's response, as not sufficiently aggressive compared to views on the same topic by Republicans. For example, the average Republican indicates a score of 3.15 relating to the Federal government response. In contrast, Democrats score 4.54, and Independents score 4.04. Republicans, compared to individuals who indicate affiliation with the other two groups, view the Federal government as acting appropriately.[29]

### B. Predictors of Social Distancing

To document the associations between individual characteristics and health behaviors, we estimate the following regression:

$$H_i = a + X_i \beta + \theta_i + \varepsilon_i$$

where $H_i$ is a measure of individual $i$'s social distancing behavior, $X_i$ is a vector of individual characteristics, $a$ is a constant term, $\theta_i$ is the county of residence fixed effects, and $\varepsilon$ is the error term. We note changes to the set of individual characteristics included in the specifications as we present the results.[30]

*Demographic Factors*. Table 6 reports the relationships between individual characteristics and various proxies of social distancing behaviors, based on the baseline data. Column (5) reports the results based on the aggregated social distancing index (higher values on the index indicate a lower tendency to social distance). We focus on the findings based on the aggregate social distancing index, as it comprises the four distinct proxies of social distancing (staying at home, non-essential outings, days outside the home in the past seven days, and the tendency to maintain at least six feet of distance from others). Columns (6)-(7) report the results on the two proxies for mask-wearing.

---

[29] These findings are consistent with polls conducted at the same time as our baseline by Pew Research, reporting that 77 percent of Republicans respond favorably to Trump's responses to the outbreak. In contrast, only 11 percent of Democrats approve of Trump's responses (Pew Research Center, 2020).
[30] We estimate linear models for continuous variables. For dichotomous variables, we estimate linear probability models.

[Insert Table 6]

The results indicate the critical role that individual demographic characteristics play in influencing social distancing. Of all the variables included in our analyses, demographic characteristics exert the most considerable influence on social distancing measures. Being Caucasian, being a Republican, and a higher household income are statistically significant predictors of and are negatively associated with social distancing. In contrast, being currently married is associated with higher rates of social distancing. In terms of effect sizes, it is noteworthy that ethnicity, political affiliation, and household income exert the most substantial influence on social distancing measures. For example, Caucasians are considerably less likely to socially distance themselves and wear masks than other ethnicities (significant at the one percent level). Republicans report lower rates of social distancing, a result significant at the five percent level, as compared to the group of individuals reporting non-Republican affiliation.

Marital status and household income are significant correlates of social distancing. Being married is associated with a higher likelihood for individuals to both have fewer non-essential outings (significant at the five percent level) and wearing a mask when shopping (significant at the five percent level). This pattern is likely related to lower demand for social contacts as married people already have more social contacts due to having a partner present within the home. Unmarried individuals may feel a greater inclination to leave home to get interaction not received inside the household. We find that those who have higher earnings are more likely to leave the house more frequently and participate in non-essential outings (significant at the five percent and one percent level, respectively).

Demographic characteristics also influence mask-wearing. Age and being female exert a positive effect on the tendency to wear a mask. Being Caucasian is negatively associated with the tendency to wear a mask. When we compare the results in Columns (5) and (7), we note a pattern: groups (ethnicity, gender, marital status, and political affiliation) that report higher mask-wearing also report lower social distancing tendencies.

We do not detect a consistent effect of educational attainment on social distancing or mask-wearing. The continuously defined education variable negatively affects mask-wearing. However, the binary indicator for a college education has a positive effect on mask-wearing (imprecisely estimated) and a positive effect on mask-wearing when shopping (significant at the five percent level).[31] We do not detect any statistically significant effect of the cognitive performance task on the social-distancing measures or mask-wearing.[32]

*Economic Preference Parameters*. Among the economic preference factors, we find that risk aversion is a strong predictor of social distancing and mask-wearing. The findings in Column (5) of Table 6 show that more risk-averse people socially distance more. Specifically, more risk-averse people are also less likely to leave the house less and report more days outside the home (both are significant at the one percent level). This pattern is consistent with recent economic studies, which also document that higher risk aversion reduces the incidence of risky health behaviors, such as cigarette smoking, heavy drinking, obesity, and seat-belt use (Anderson and Miller, 2008; de Oliveira et al., 2016).

We find no evidence of time preference influencing social distancing measures. However, we find a surprising relationship: people with lower willingness to delay gratification (i.e., higher impatience) exhibit a higher tendency to wear a mask, controlling for other factors. A similar pattern is documented by Scharff and Viscusi (2011) in the context of time discounting and smoking status. Scharff & Viscusi (2011) find that current smokers have a higher implied rate of time preference than nonsmokers. Chapman (2005) argues that the strength and the direction of the association between better health behaviors and time discounting depend critically on the specific preventive health behavior.[33]

We find strong evidence that personality traits are significant predictors of social distancing measures. The three most crucial traits for more social distancing are higher positive reciprocity, higher openness, and being more agreeable.[34]

---

[31] Cutler and Lleras-Muney (2010) find a large and persistent positive association between education and health. They find that the positive effect of education increases with increasing years of education, with no evidence of a sheepskin effect.
[32] Bijwaard et al. (2016) examine the importance of cognitive skills for health outcomes. The study finds that for most ages, cognitive ability explains around half of the raw differences in health outcomes (mortality) across educational groups.
[33] Based on a meta-analysis study, Chapman (2005) shows that 'hot' or addictive health behaviors such as smoking, are more likely to associated with time preference than 'cold' behaviors such as vaccination. Other prevention behaviors designed to prevent cervical and breast cancer are associated with individuals with higher life expectancy, lower time preference, and more risk aversion (Picone et al. 2004).
[34] Bogg & Roberts (2004) find that individuals who are conscientious engage in less risky health-behaviors. Other studies (Post, 2005; Borgonovi, 2008) also find a positive correlation between higher altruism and better health outcomes. Bessey (2018) finds that agreeableness is negatively related to the probability of drinking and engaging in physical activity. Brummett et al., (2006) finds that personality traits are a significant predictor of BMI; the study finds that extraversion, neuroticism, and conscientiousness determine weight gain in adulthood.

***The Role of Misperception***. We examine the possibility that exponential growth bias (EGB) could influence social distancing behavior. Specifically, we regress the proxies of social distancing on the exponential growth bias index, controlling for all other socio-economic characteristics and economic preference factors. On the aggregate social distancing index, people who misunderstand exponential growth (i.e., exhibit higher exponential growth bias) report a lower tendency to social distance. Specifically, higher EGB strongly predicts the number of days one leaves the house per week. EGB influences the other proxies of social distancing (in Table 6), but the effect size estimates are imprecise. Given the importance of EGB in predicting social distancing outcomes, we return to examining what factors predict the presence of EGB itself.

We collect detailed information on the CFR (as defined in Section II.C) associated with COVID-19 at the county level. Therefore, we can conduct additional analyses on people's subjective assessments of the disease risk associated with this dimension of the disease. The average CFR for counties associated with the place of residence for people in our sample and at the time of the baseline (i.e., the last week of May 2020) is 0.06. This rate mirrors the average rate for the United States during the last week of May (European Centre for Disease Prevention and Control, 2020).

Using data from the panel survey, we examine the accuracy of the respondents' assessment regarding the CFR in their county of residence. Despite the limitations of the measures of the subjective assessment of the CFR, it is worth exploring whether they predict social distancing decisions. The average subjective assessment of the CFR, associated with the respondent's county of residence in our sample, is 0.07 (slightly higher than the average objective county-level CFR for the counties of the individuals in our sample). In contrast, the median subjective assessment is 0.03. These numbers imply that a considerable number of respondents underestimate the risks associated with COVID-19 (67 percent of the sample reports a subjective assessment that is lower than the objective CFR in their county of residence) and that a small fraction overstates the risks by a lot.

We examine how the accuracy of risk perception associated with COVID-19 predicts social distancing behavior. Table 6 reports the results. We do not detect statistically significant effects on the aggregate social distancing index, based on the percent definition of the risk misperception variable. Based on the same specifications, we also conduct additional analyses based on binary definitions for subgroups with wildly inaccurate misperceptions (i.e., more than 10 percent difference from the objective CFR). Results, based on these subgroups, are suggestive of negative effects on the aggregate social distancing index, but they are not statistically significant at conventional levels. Table 6 (Column 6) also reports the results for the mask-wearing outcome: the results show that individuals who misperceive the CFR risk associated with COVID-19 are considerably less likely to wear a mask.

***Predictors to Changes of Own Behavior***. Because we rely on longitudinal data from the same people, we can also examine whether any variables predict changes to individual behavior. An important caveat to this analysis is that it relies on a limited time-period (three weeks) and fewer observations due to some panel attrition. On the other hand, examining longitudinal differences can be useful for policies that aim at influencing changes to behavior. Table 7 reports the findings of this analysis.

[Insert Table 7]

We find that six factors are significant predictors of changes to one's behavior regarding social distancing. Specifically, being currently married, being Caucasian, intense COVID-19 fears[35], living in a more densely populated county, higher cognitive performance, and being more agreeable[36] (the Big Five index) are important drivers of changes in social distancing outcomes.

### C. Determinants of Risk Misperception and Exponential Growth Bias

---

[35] We constructed a COVID-19 fear indicator as a binary measure. The variable is based on the respondent indicating that he/she worries on daily to the following question: "How frequently have you feared being infected with COVID-19 in the past month?"

[36] This finding is consistent with a meta-analysis study, conducted by Kline et al. (2019), showing that agreeableness is a consistent and significant predictor of increased prosocial behavior.

As we noted earlier, the two perception variables, risk misperception and EGB, predict social distancing or mask-wearing. Therefore, we examine the drivers of the EGB and risk misperception. A better understanding of the predictors of these two phenomena will shed light on mechanisms that will help the design of policies geared towards reducing EGB and risk misperception. Tables 8 and 9 report the findings.

The strongest predictors of individual misperceptions relate to gender, ethnicity, political affiliation, population density, and concerns about their health. We find that females are more likely to overstate the risk of COVID-19 (related to the CFR measure defined in Section II). Caucasians and people who reported Republican affiliation exhibit lower risk perception (significant at the one percent level). The same two demographic groups are not only the ones reporting lower subjective risk perception, but also the ones considerably understating the COVID-19 risk in areas where the COVID-19 risks (associated with the CFR) are objectively much higher than the subjective assessments.

[Insert Table 8] [Insert Table 9]

Exponential growth bias is best predicted by age, gender, schooling, cognitive skills, and neuroticism. In terms of the effect size estimates related to the predictors, gender, cognitive score, and schooling exert the largest effect on the EGB. Older individuals, being female, and the neuroticism trait positively predict EGB. In contrast, higher educational attainment, better cognitive performance, and higher earnings considerably reduce the magnitude of EGB.[37] Although not reported here, a related online field experiment examines the role informational provision can play in influencing EGB and subsequent behaviors.

**IV. Concluding Remarks**

Understanding the determinants of what influences preventive measures against the COVID-19 virus is paramount for effective policies to be successful in encouraging people to adopt social distancing measures and mask-wearing. Absent a vaccine, the best we can do is structure policy aimed at those failing to adhere to preventive health measures to curb the spread of COVID-19. However, if people do not have the incentives to do so on their own, the implementation of policies or mandates are unlikely to succeed. To this end, we note several important lessons from the preliminary analysis based on the panel data from participants in our study.

---

[37] It is noteworthy that the EGB measure is also highly correlated with higher risk aversion and higher patience (time preference index).

First, the key drivers of social distancing measures are demographic ones, even when we control for numerous other factors. Of all the variables included in our analyses, demographic characteristics exert the most considerable influence on social distancing measures. Being Caucasian, being a Republican, and higher household income are statistically significant predictors of lower rates of social distancing. In contrast, being currently married is positively associated with higher rates of social distancing.

The importance of demographic variables for influencing social distancing is somewhat puzzling. They may account for critical, unobservable characteristics that are hard to measure. For example, the demographic characteristics may pick up mechanisms related to how well-informed individual decision-makers are, or how information about the virus, the risks associated with, or about the effectiveness of social distancing measures gets shared across social networks. On the other hand, it is also possible that the decision to adopt social distancing is much more related to demographic characteristics or political affiliation due to the enormous politicization of the issue.

Second, and based on the longitudinal data collected between the follow-up surveys and the baseline data, it seems that demographic factors are also some of the factors that can explain changes among individuals who reported changes in their behaviors. Based on our analyses, females, currently married people, and people living in densely populated areas are especially prone to change their social distancing behaviors. Knowledge of what demographic groups are more prone to making changes to their social distancing and mask-wearing can be enormously beneficial from a policy targeting perceptive.

Finally, we document the importance of two factors that are especially prone to information barriers, and that can also influence social distancing behaviors: risk misperception and exponential growth bias. Highlighting the role these factors play on influencing behaviors underscores the potential use for policies geared towards correcting these perception biases.

Policies that target demographic groups known to understate the risks associated with COVID-19 or misinterpreting the speed at which it spreads can be effective.[38] The findings from this study are informative for crafting such policies. Notwithstanding some limitations, the present study offers additional direct evidence on what demographic groups could be particularly suitable for such interventions.

---

[38] Lammers et al. (2020) discuss the effectiveness of such policies.

**Table 1.** Summary Statistics

|  | Full Sample | Women | Men |
|---|---|---|---|
|  | (1) | (2) | (3) |
| **Socioeconomic Characteristics:** | | | |
| Age | 42.40 (19.14) | 44.88 (18.83) | 32.99 (17.36) |
| Years of Schooling | 15.35 (2.42) | 15.53 (2.37) | 14.68 (2.46) |
| Four-Year College Graduate (=1 if yes) | 0.61 (0.49) | 0.65 (0.48) | 0.46 (0.50) |
| Current Full-time Student (=1 if yes) | 0.09 (0.28) | 0.08 (0.27) | 0.12 (0.33) |
| Caucasian (=1 if yes) | 0.80 (0.40) | 0.82 (0.38) | 0.70 (0.46) |
| African American (=1 if yes) | 0.02 (0.14) | 0.02 (0.14) | 0.03 (0.16) |
| Female (=1 if yes) | 0.79 (0.41) | - | - |
| Currently Married (=1 if yes) | 0.41 (0.49) | 0.45 (0.50) | 0.24 (0.43) |
| Republican (=1 if yes) | 0.11 (0.32) | 0.11 (0.31) | 0.13 (0.33) |
| Democrat (=1 if yes) | 0.60 (0.49) | 0.61 (0.49) | 0.55 (0.50) |
| Currently Employed (=1 if yes) | 0.48 (0.50) | 0.50 (0.50) | 0.42 (0.49) |
| Household Annual Income (in USD) | 115924.00 (75520.69) | 115701.30 (74505.39) | 116768.60 (79453.96) |
| Health Status [a] | 4.04 (0.78) | 4.02 (0.79) | 4.11 (0.75) |
| Ever smoked before outbreak (=1 if yes) | 0.27 (0.44) | 0.28 (0.45) | 0.24 (0.43) |
| Current smoker (=1 if yes) | 0.05 (0.23) | 0.06 (0.24) | 0.04 (0.19) |
| Alcohol frequency (before outbreak) [b] | 1.61 (1.80) | 1.59 (1.80) | 1.67 (1.79) |
| Alcohol currently [b] | 1.75 (2.13) | 1.78 (2.16) | 1.68 (2.04) |
| Concern about own health [c] | 4.29 (2.65) | 4.39 (2.68) | 3.89 (2.52) |
| **COVID-19-related Characteristics:** | | | |
| Likelihood contracting COVID-19 (percent) | 0.34 (0.26) | 0.35 (0.26) | 0.30 (0.27) |
| Chance of severe symptoms after COVID-19 infection (percent) | 0.48 (0.30) | 0.51 (0.30) | 0.39 (0.30) |
| Subjective Assessment Case Fatality Rate of COVID-19 (percent) | 0.08 (0.12) | 0.07 (0.12) | 0.08 (0.13) |
| Subjective Assessment Case Fatality Rate of COVID-19 in own demographic category (percent) | 0.05 (0.11) | 0.06 (0.11) | 0.04 (0.09) |
| County of Residence Case Fatality Rate (percent) | 0.06 (0.03) | 0.06 (0.03) | 0.06 (0.03) |
| **Other Characteristics:** | | | |
| Cognitive Score [d] | 0.94 (0.48) | 0.90 (0.45) | 1.09 (0.54) |
| Time Preferences Index [e] | 20.61 (10.43) | 20.63 (10.26) | 20.56 (11.06) |
| Risk-Preference Index [f] | 5.10 (2.97) | 5.01 (3.04) | 5.47 (2.69) |
| Patient (=1 if yes) | 0.23 (0.42) | 0.21 (0.41) | 0.27 (0.45) |
| Impatient (=1 if yes) | 0.11 (0.31) | 0.10 (0.30) | 0.12 (0.33) |
| Time-Inconsistent (=1 if yes) [g] | 0.25 (0.44) | 0.25 (0.43) | 0.27 (0.45) |
| Observations | 901 | 713 | 188 |

*Notes*: Standard deviations are reported in parenthesis. Behavior outcomes based on baseline wave of survey. (a) Health Status: 1 being "very poor" and 5 being "excellent" (b) Alcohol consumption is the number of days per week in which the respondent reported having a drink (c) Worry about own health: 0=not at all, 10=a lot. (d) Cognitive score is derived based on total possible correct answers (=4) on a Raven's Matrices test divided by the time allotted to task (two minutes). (e) Time preference index from 1 (extremely impatient) to 32 (extremely patient) based on a near-frame question. Impatient (=1 if time preference index is 1). Patient (=1 if time preference index is 32). (f) Risk preference index from 0 (extremely risk tolerant) to 10 (extremely risk averse). (g) "Time-inconsistent" is defined with respect to two "money" questions: 1) a near-frame question prompting individuals to choose between 160 USD today and a larger amount of USD in 12 months. 2) a distant-frame question prompting individuals to choose between 160 USD in 12 months and a larger amount of USD in 24 months. Individuals were identified as "hyperbolic" time-inconsistent if they preferred the immediate reward in the near-term frame combined with the choice of the delayed reward in the distant frame.

Table 2. Summary by Study States

|  | CT | FL | MA | MD | NJ | NY | PA |
|---|---|---|---|---|---|---|---|
|  | (1) | (2) | (3) | (4) | (5) | (6) | (7) |
| **Socioeconomic Characteristics:** | | | | | | | |
| Age | 50.83 (16.06) | 60.29 (11.41) | 52.51 (15.76) | 52.17 (14.61) | 47.04 (16.16) | 28.29 (14.13) | 53.70 (13.67) |
| Years of Schooling | 16.31 (2.47) | 16.14 (2.05) | 16.51 (2.21) | 16.47 (1.74) | 16.15 (2.46) | 14.05 (2.04) | 16.57 (2.17) |
| Four-Year College Graduate (=1 if yes) | 0.77 (0.42) | 0.79 (0.41) | 0.82 (0.39) | 0.86 (0.35) | 0.75 (0.44) | 0.36 (0.48) | 0.82 (0.38) |
| Current Full-time Student (=1 if yes) | 0.08 (0.28) | - | - | 0.02 (0.13) | 0.03 (0.16) | 0.17 (0.38) | 0.02 (0.14) |
| Caucasian (=1 if yes) | 0.92 (0.28) | 0.95 (0.21) | 0.94 (0.24) | 0.91 (0.28) | 0.84 (0.37) | 0.62 (0.49) | 0.97 (0.16) |
| African American (=1 if yes) | 0.02 (0.14) | - | 0.01 (0.12) | 0.02 (0.13) | 0.01 (0.11) | 0.04 (0.19) | - |
| Female (=1 if yes) | 0.83 (0.38) | 0.85 (0.36) | 0.90 (0.31) | 0.90 (0.31) | 0.88 (0.33) | 0.70 (0.46) | 0.84 (0.36) |
| Currently Married (=1 if yes) | 0.46 (0.50) | 0.68 (0.47) | 0.55 (0.50) | 0.66 (0.48) | 0.59 (0.50) | 0.13 (0.34) | 0.67 (0.47) |
| Republican (=1 if yes) | 0.10 (0.31) | 0.18 (0.39) | - | 0.16 (0.37) | 0.08 (0.27) | 0.09 (0.29) | 0.18 (0.39) |
| Democrat (=1 if yes) | 0.69 (0.47) | 0.54 (0.50) | 0.67 (0.47) | 0.67 (0.47) | 0.68 (0.47) | 0.56 (0.50) | 0.59 (0.49) |
| Currently Employed (=1 if yes) | 0.54 (0.50) | 0.48 (0.50) | 0.64 (0.48) | 0.71 (0.46) | 0.58 (0.50) | 0.35 (0.48) | 0.61 (0.49) |
| Household Annual Income (in USD) | 135572.90 (69975.75) | 114023.80 (72582.43) | 125074.60 (72156.93) | 144137.90 (68831.97) | 151062.50 (67088.85) | 99658.23 (77372.85) | 120118.20 (71450.20) |
| Health Status[a] | 4.00 (0.88) | 4.01 (0.70) | 3.97 (0.90) | 4.12 (0.75) | 3.98 (0.76) | 4.09 (0.74) | 3.99 (0.86) |
| Ever smoked before outbreak (=1 if yes) | 0.15 (0.36) | 0.38 (0.49) | 0.33 (0.47) | 0.22 (0.42) | 0.25 (0.44) | 0.24 (0.43) | 0.31 (0.46) |
| Current smoker (=1 if yes) | - | 0.10 (0.31) | 0.06 (0.24) | 0.05 (0.22) | 0.04 (0.19) | 0.05 (0.22) | 0.05 (0.21) |
| Alcohol frequency (before outbreak)[b] | 1.63 (1.92) | 2.10 (2.31) | 1.93 (2.12) | 1.55 (1.63) | 1.56 (1.73) | 1.43 (1.51) | 1.65 (1.96) |
| Alcohol currently[b] | 1.92 (2.24) | 2.39 (2.49) | 2.15 (2.35) | 2.29 (2.24) | 1.89 (2.01) | 1.27 (1.79) | 2.08 (2.34) |
| Concern about own health[c] | 4.79 (2.58) | 4.87 (2.77) | 4.58 (3.07) | 4.28 (2.66) | 4.76 (2.61) | 3.88 (2.45) | 4.43 (2.77) |
| **Respondents Beliefs Regarding COVID-19:** | | | | | | | |
| Likelihood contracting COVID-19 (percent) | 0.34 (0.23) | 0.31 (0.24) | 0.37 (0.32) | 0.36 (0.26) | 0.33 (0.26) | 0.34 (0.27) | 0.32 (0.25) |
| Chance of severe symptoms after COVID-19 infection (percent) | 0.54 (0.25) | 0.60 (0.31) | 0.55 (0.33) | 0.47 (0.24) | 0.52 (0.31) | 0.42 (0.29) | 0.52 (0.30) |
| Subjective Assessment Case Fatality Rate of COVID-19 (percent) | 0.05 (0.07) | 0.04 (0.06) | 0.06 (0.10) | 0.04 (0.08) | 0.07 (0.13) | 0.11 (0.15) | 0.04 (0.07) |
| Subjective Assessment Case Fatality Rate of COVID-19 in own demographic category (percent) | 0.03 (0.04) | 0.06 (0.10) | 0.07 (0.14) | 0.05 (0.14) | 0.05 (0.10) | 0.05 (0.10) | 0.06 (0.12) |
| **Other Characteristics:** | | | | | | | |
| Cognitive Score[d] | 0.81 (0.48) | 0.88 (0.44) | 0.94 (0.47) | 0.87 (0.46) | 0.92 (0.42) | 1.01 (0.51) | 0.86 (0.41) |
| Time Preferences Index[e] | 22.66 (9.87) | 20.53 (10.13) | 22.75 (9.09) | 22.59 (11.04) | 21.49 (10.07) | 19.09 (10.85) | 21.86 (9.70) |
| Risk-Preference Index[f] | 4.38 (3.27) | 5.06 (2.90) | 4.94 (2.90) | 5.42 (2.67) | 4.80 (3.29) | 5.20 (2.95) | 5.24 (3.00) |
| Patient (=1 if yes) | 0.31 (0.47) | 0.21 (0.41) | 0.18 (0.39) | 0.40 (0.49) | 0.20 (0.40) | 0.19 (0.40) | 0.26 (0.44) |
| Impatient (=1 if yes) | 0.08 (0.28) | 0.09 (0.28) | 0.06 (0.24) | 0.10 (0.31) | 0.10 (0.30) | 0.13 (0.34) | 0.08 (0.27) |
| Time-Inconsistent (=1 if yes)[g] | 0.19 (0.39) | 0.14 (0.35) | 0.28 (0.45) | 0.19 (0.40) | 0.30 (0.46) | 0.28 (0.45) | 0.26 (0.44) |
| Observations | 48 | 105 | 67 | 58 | 80 | 395 | 148 |

*Notes*: Standard deviations are reported in parenthesis. Behavior outcomes based on baseline wave of survey. (a) Health Status: 1 being "very poor" and 5 being "excellent" (b) Alcohol consumption is the number of days per week in which the respondent reported having a drink (c) Worry about own health: 0=not at all, 10=a lot. (d) Cognitive score is derived based on total possible correct answers (=4) on a Raven's Matrices test divided by the time allotted to task (two minutes). (e) Time preference index from 1 (extremely impatient) to 32 (extremely patient) based on a near-frame question. Impatient (=1 if time preference index is 1). Patient (=1 if time preference index is 32). (f) Risk preference index from 0 (extremely risk tolerant) to 10 (extremely risk averse). (g) "Time-inconsistent" is defined with respect to two "money" questions: 1) a near-frame question prompting individuals to choose between 160 USD today and a larger amount of USD in 12 months. 2) a distant-frame question prompting individuals to choose between 160 USD in 12 months and a larger amount of USD in 24 months. Individuals were identified as "hyperbolic" time-inconsistent if they preferred the immediate reward in the near-term frame combined with the choice of the delayed reward in the distant frame.

**Table 3.** Summary Social Distancing by Demographic Characteristics

| | Social Distancing Behavior (Baseline) | | | | | Social Distancing Behavior (Change)[a] | | | |
|---|---|---|---|---|---|---|---|---|---|
| | Did not Maintain Six Feet (=1 if yes) | Did not Stay Home (=1 if yes) | Non-essential Outings (#) | Days Out Past Week (#) | Social Distancing Index[b] | Did not Maintain Six Feet | Did not Stay Home | Days Out Past Week (#) | Social Distancing Index[b] |
| | (1) | (2) | (3) | (4) | (5) | (6) | (7) | (9) | (10) |
| All | 0.47 (0.47) | 0.70 (0.46) | 0.74 (1.03) | 3.03 (2.10) | 0.00 (1.33) | 0.17 (0.55) | 0.16 (0.51) | 0.49 (1.70) | 8.09e-09 (1.89) |
| High School Diploma (=1 if yes) | 0.51 (0.50) | 0.69 (0.47) | 1.08 (1.20) | 2.64 (1.85) | 0.01 (1.32) | 0.13 (0.54) | 0.11 (0.55) | 0.35 (1.83) | -0.14 (1.34) |
| Some College (=1 if yes) | 0.52 (0.50) | 0.74 (0.44) | 0.98 (1.16) | 2.88 (2.07) | 0.13 (1.33) | 0.12 (0.52) | 0.12 (0.49) | 0.47 (1.65) | -0.09 (1.17) |
| College Degree (=1 if yes) | 0.44 (0.50) | 0.68 (0.46) | 0.56 (0.88) | 3.19 (2.15) | -0.06 (1.34) | 0.19 (0.56) | 0.19 (0.50) | 0.53 (1.69) | 0.07 (1.15) |
| Graduate School (=1 if yes) | 0.40 (0.49) | 0.62 (0.49) | 0.51 (0.84) | 3.04 (2.23) | -0.23 (1.43) | 0.20 (0.53) | 0.17 (0.50) | 0.65 (1.85) | 0.10 (1.19) |
| Democrat (=1 if yes) | 0.43 (0.49) | 0.69 (0.46) | 0.66 (0.95) | 3.03 (2.11) | -0.07 (1.28) | 0.17 (0.55) | 0.15 (0.49) | 0.46 (1.67) | -0.03 (1.13) |
| Republican (=1 if yes) | 0.53 (0.50) | 0.81 (0.39) | 0.90 (1.19) | 3.21 (2.06) | 0.25 (0.43) | 0.24 (0.47) | 0.17 (0.48) | 0.67 (1.94) | 0.14 (1.19) |
| Independent or Other (=1 if yes) | 0.52 (0.50) | 0.68 (0.47) | 0.85 (1.11) | 2.97 (2.11) | 0.04 (1.40) | 0.13 (0.57) | 0.19 (0.55) | 0.49 (1.68) | 0.01 (1.29) |
| Caucasian (=1 if yes) | 0.48 (0.50) | 0.73 (0.44) | 0.69 (0.99) | 3.22 (2.09) | 0.07 (1.31) | 0.17 (0.55) | 0.16 (0.49) | 0.53 (1.66) | 0.02 (1.16) |
| African American (=1 if yes) | 0.53 (0.51) | 0.74 (0.45) | 1.16 (1.30) | 0.21 (0.42) | 0.08 (1.69) | 0.25 (0.62) | 0.00 (0.00) | 0.17 (1.70) | -0.24 (0.85) |
| Hispanic (=1 if yes) | 0.43 (0.50) | 0.64 (0.49) | 1.11 (1.20) | 2.93 (2.32) | -0.04 (1.49) | 0.19 (0.40) | 0.19 (0.56) | 0.07 (1.99) | -0.09 (1.38) |
| Other Ethnicity (=1 if yes) | 0.44 (0.50) | 0.55 (0.50) | 0.83 (1.11) | 2.14 (1.82) | -0.42 (1.25) | 0.12 (0.61) | 0.18 (0.59) | 0.43 (1.84) | -0.03 (1.34) |
| Observations | 901 | 901 | 901 | 901 | 901 | 623 | 623 | 623 | 623 |

*Notes*: Education levels: Graduate school includes the following degrees: MD, DDS, DVM, LLB, JD, and Doctorate Degrees (ex: PhD, EdD). (a) Represents the change in behavior from the baseline to the week two follow-up survey. (b) The social distancing index is based on a PCA procedure based on the four distinct proxies (did not maintain six feet in previous week; did not stay home in previous week; number of non-essential outings; days out of home in previous week) of social distancing.

**Table 4.** Summary Mask Wearing by Demographic Characteristics

|  | Wears a Mask (=1 if yes) | Wears a Mask when Shopping (=1 if yes) | Wears a Mask (Change) [a] |
|---|---|---|---|
|  | (1) | (2) | (3) |
| All | 0.68 (0.47) | 0.91 (0.28) | -0.01 (0.49) |
| High School Diploma (=1 if yes) | 0.64 (0.48) | 0.86 (0.35) | 0.00 (0.47) |
| Some College (=1 if yes) | 0.67 (0.47) | 0.89 (0.31) | -0.05 (0.53) |
| College Degree (=1 if yes) | 0.70 (0.46) | 0.93 (0.25) | 0.00 (0.47) |
| Graduate School (=1 if yes) | 0.61 (0.49) | 0.88 (0.33) | -0.06 (0.53) |
| Democrat (=1 if yes) | 0.71 (0.46) | 0.91 (0.28) | -0.02 (0.50) |
| Republican (=1 if yes) | 0.63 (0.49) | 0.87 (0.34) | 0.02 (0.54) |
| Independent or Other (=1 if yes) | 0.65 (0.48) | 0.93 (0.26) | 0.01 (0.46) |
| Caucasian (=1 if yes) | 0.67 (0.47) | 0.92 (0.27) | 0.01 (0.48) |
| African American (=1 if yes) | 0.63 (0.50) | 0.84 (0.37) | 0.17 (0.58) |
| Hispanic (=1 if yes) | 0.73 (0.45) | 0.93 (0.25) | -0.15 (0.53) |
| Other Ethnicity (=1 if yes) | 0.71 (0.45) | 0.88 (0.32) | -0.06 (0.53) |
| Observations | 901 | 901 | 623 |

*Notes*: Graduate school includes the following degrees: MD, DDS, DVM, LLB, JD, and Doctorate Degrees (ex: PhD, EdD). (a) Represents the change in behavior from the baseline to the week two follow-up survey.

Table 5. Summary Opinion of Government Response by Political Affiliation

|  | Federal Government Reaction | State Government Reaction | Local Government Reaction | General Public Reaction |
|---|---|---|---|---|
|  | (1) | (2) | (3) | (4) |
| All | 4.24 (0.94) | 3.22 (0.78) | 3.29 (0.75) | 3.77 (1.00) |
| Republican (=1 if yes) | 3.15 (1.02) | 2.74 (1.08) | 2.86 (0.91) | 3.14 (1.21) |
| Democrat (=1 if yes) | 4.54 (0.68) | 3.35 (0.66) | 3.40 (0.64) | 3.95 (0.88) |
| Independent and Other (=1 if yes) | 4.04 (0.99) | 3.15 (0.78) | 3.24 (0.83) | 3.64 (1.04) |
|  |  |  |  |  |
| Observations | 892 | 898 | 849 | 885 |

*Notes:* The attitude questions were scale-based, using scale the following scale: 1=reaction is much too extreme to 5=reaction is not at all sufficient. The questions are: reaction to federal government's response ("Do you think the reaction of the federal government to the current coronavirus outbreak is appropriate, too extreme, or not sufficient?"), reaction to state government's response ("Do you think the reaction of your state government to the current coronavirus outbreak is appropriate, too extreme, or not sufficient?"), reaction to local government's response ("Do you think the reaction of your local government to the current coronavirus outbreak is appropriate, too extreme, or not sufficient?"), reaction to the general public's response ("Do you think the reaction of most people in the United States is appropriate, too extreme, or not sufficient?").

**Table 6.** Determinants of Proxy Measures of Social Distancing

| | Did not Stay Home (=1 if yes) | Non-essential Outings (#) | Days out past week (#) | Did not Maintain Six Feet (=1 if yes) | Social Distancing Index[a] | Wears a Mask (=1 if yes) | Wears a Mask when Shopping (=1 if yes) |
|---|---|---|---|---|---|---|---|
| | (1) | (2) | (3) | (4) | (5) | (6) | (7) |
| Age | 0.00* | -0.00** | 0.01 | -0.00* | -0.00 | 0.00*** | 0.00* |
| | (0.00) | (0.00) | (0.01) | (0.00) | (0.00) | (0.00) | (0.00) |
| Female (=1 if yes) | 0.00 | -0.20** | -0.21 | 0.05 | -0.09 | 0.12*** | 0.03 |
| | (0.04) | (0.09) | (0.18) | (0.04) | (0.11) | (0.04) | (0.02) |
| Currently Married (=1 if yes) | -0.03 | -0.18** | -0.23 | -0.07 | -0.20* | 0.04 | 0.05** |
| | (0.04) | (0.08) | (0.18) | (0.04) | (0.11) | (0.04) | (0.02) |
| Caucasian (=1 if yes) | 0.13*** | 0.06 | 0.41** | 0.09* | 0.38*** | -0.12*** | 0.02 |
| | (0.04) | (0.10) | (0.21) | (0.05) | (0.13) | (0.04) | (0.03) |
| Schooling | -0.02* | 0.01 | -0.03 | -0.01 | -0.04 | -0.02* | -0.01 |
| | (0.01) | (0.03) | (0.06) | (0.01) | (0.04) | (0.01) | (0.01) |
| College Graduate (=1 if yes) | 0.01 | -0.19 | 0.14 | -0.01 | -0.02 | 0.07 | 0.08** |
| | (0.06) | (0.13) | (0.27) | (0.07) | (0.17) | (0.06) | (0.03) |
| Republican (=1 if yes) | 0.10** | 0.26** | 0.04 | 0.05 | 0.28** | -0.05 | -0.05 |
| | (0.05) | (0.11) | (0.23) | (0.05) | (0.14) | (0.04) | (0.03) |
| Earnings (Logged) | 0.01 | 0.13*** | 0.18** | 0.03 | 0.11** | -0.01 | -0.00 |
| | (0.02) | (0.04) | (0.08) | (0.02) | (0.05) | (0.02) | (0.01) |
| Population Density (Logged) | -0.01 | -0.00 | -0.07** | -0.01* | -0.05** | 0.01 | 0.00 |
| | (0.01) | (0.02) | (0.04) | (0.01) | (0.02) | (0.01) | (0.01) |
| Concerned About Health[b] | -0.01** | 0.01 | -0.06** | -0.00 | -0.03** | 0.01* | 0.01* |
| | (0.01) | (0.02) | (0.03) | (0.01) | (0.02) | (0.01) | (0.00) |
| COVID19 Infection Fears on a Daily Basis[c] | -0.04 | -0.04 | 0.37** | -0.04 | -0.00 | 0.05 | -0.01 |
| | (0.04) | (0.09) | (0.18) | (0.04) | (0.11) | (0.04) | (0.02) |
| Number Comorbidities | 0.02 | -0.12* | -0.36** | -0.00 | -0.11 | 0.04 | 0.01 |
| | (0.03) | (0.07) | (0.15) | (0.03) | (0.09) | (0.03) | (0.02) |
| Cognitive Score[d] | 0.00 | -0.03 | -0.05 | 0.01 | -0.01 | -0.01 | 0.02 |
| | (0.02) | (0.04) | (0.07) | (0.02) | (0.05) | (0.02) | (0.01) |
| Risk Aversion Index[e] | -0.02*** | -0.02 | -0.05** | -0.00 | -0.05*** | 0.00 | 0.00 |
| | (0.01) | (0.01) | (0.03) | (0.01) | (0.02) | (0.01) | (0.00) |
| Time Preference Index[f] | 0.00 | 0.00 | 0.00 | 0.00* | 0.01 | -0.00** | 0.00 |
| | (0.00) | (0.00) | (0.01) | (0.00) | (0.00) | (0.00) | (0.00) |
| Time Inconsistent (=1 if yes)[g] | 0.04 | 0.06 | 0.04 | -0.00 | 0.09 | -0.04 | 0.01 |
| | (0.04) | (0.08) | (0.18) | (0.04) | (0.11) | (0.04) | (0.02) |
| Positive Reciprocity Index | -0.01 | -0.04* | -0.07 | -0.01 | -0.05* | 0.02** | -0.00 |
| | (0.01) | (0.02) | (0.04) | (0.01) | (0.03) | (0.01) | (0.01) |
| Altruism Index | -0.01 | 0.00 | 0.00 | -0.01 | -0.03 | 0.02** | 0.00 |
| | (0.01) | (0.02) | (0.04) | (0.01) | (0.03) | (0.01) | (0.01) |

(Continued on next page)

**Table 6 (Continued):** Determinants of Social Distancing

|  | Did not Stay Home (=1 if yes) | Non-essential Outings (#) | Days Out Past Week (#) | Did not Maintain Six Feet (=1 if yes) | Social Distancing Index | Wears a Mask (=1 if yes) | Wears a Mask when Shopping (=1 if yes) |
|---|---|---|---|---|---|---|---|
|  | (1) | (2) | (3) | (4) | (5) | (6) | (7) |
| Exponential Growth Bias Index[h] | 0.01 | 0.01 | 0.11** | 0.01 | 0.05* | 0.01 | -0.01* |
|  | (0.02) | (0.02) | (0.04) | (0.01) | (0.02) | (0.01) | (0.01) |
| Risk Misperception (in percent)[i] | 0.13 | 0.39 | -0.13 | 0.07 | 0.33 | -0.26** | -0.03 |
|  | (0.13) | (0.28) | (0.60) | (0.14) | (0.38) | (0.13) | (0.08) |
| Openness Index[j] | 0.03 | 0.03 | 0.16** | 0.00 | 0.09* | 0.03* | 0.02* |
|  | (0.02) | (0.04) | (0.07) | (0.02) | (0.05) | (0.02) | (0.01) |
| Extraversion Index[j] | -0.00 | 0.01 | -0.03 | -0.00 | -0.01 | 0.02 | -0.01 |
|  | (0.02) | (0.03) | (0.07) | (0.02) | (0.05) | (0.02) | (0.01) |
| Conscientiousness Index[j] | -0.01 | 0.02 | -0.03 | -0.02 | -0.04 | 0.03* | -0.01 |
|  | (0.02) | (0.04) | (0.07) | (0.02) | (0.05) | (0.02) | (0.01) |
| Agreeableness Index[j] | -0.03 | -0.08** | -0.11 | -0.04** | -0.13*** | 0.02 | 0.00 |
|  | (0.02) | (0.04) | (0.07) | (0.02) | (0.05) | (0.02) | (0.01) |
| Neuroticism Index[j] | -0.01 | -0.02 | -0.04 | -0.02 | -0.05 | 0.02 | 0.01 |
|  | (0.02) | (0.04) | (0.08) | (0.02) | (0.05) | (0.02) | (0.01) |
|  |  |  |  |  |  |  |  |
| Baseline Mean | 0.70 | 0.74 | 3.03 | 0.47 | 0.00 | 0.70 | 0.91 |
| Controls | Yes | Yes | Yes | Yes | Yes | Yes | Yes |
| R-squared | 0.05 | 0.12 | 0.06 | 0.03 | 0.06 | 0.10 | 0.02 |
|  |  |  |  |  |  |  |  |
| Observations | 901 | 901 | 901 | 901 | 901 | 901 | 901 |

*Notes:* Clustered standard errors are reported in parenthesis. All models include state and regional fixed effects. Behavior outcomes based on baseline wave of survey. (a) The social distancing index is based on a PCA procedure based on the four distinct proxies (did not maintain six feet in previous week; did not stay home in previous week; number of non-essential outings; days out of home in previous week) of social distancing. (b) Worry about own health: 0=not at all, 10=a lot. (c) COVID-19 Fear is constructed as a binary measure based on the respondent indicating that he/she worries on daily to the following question "How frequently have you feared being infected with COVID-19 in the past month?" (d) Cognitive score is derived based on total possible correct answers (=4) divided by time allocation for task (two minutes). (e) Risk preference index from 0 (extremely risk tolerant) to 10 (extremely risk averse). (f) Time preference index from 1 (extremely impatient) to 32 (extremely patient) based on a near-frame question. Impatient (=1 if time preference index is 1). Patient (=1 if time preference index is 32). (g) "Time-inconsistent" is defined with respect to two "money" questions: 1) a near-frame question prompting individuals to choose between 160 USD today and a larger amount of USD in 12 months. 2) a distant-frame question prompting individuals to choose between 160 USD in 12 months and a larger amount of USD in 24 months. Individuals were identified as "hyperbolic" time-inconsistent if they preferred the immediate reward in the near-term frame combined with the choice of the delayed reward in the distance frame. (h) The exponential growth bias index is based on the number of incorrect responses on five exponential growth-related questions. Higher values on the index indicate higher exponential growth bias. (i) Risk misperception is the difference between the subjectively elicited case fatality rate in own county minus the actual case fatality rate at the time of the interview based on data from the Centers for Disease Control and Prevention. (j) The five variables (openness, extraversion, conscientiousness, agreeableness, neuroticism) are based on a taxonomy from psychology theory for the "Big Five" personality traits. They are indices (higher values on the index indicate increased tendency of the particular trait).
***Significant at the 1 percent level.
**Significant at the 5 percent level.
*Significant at the 10 percent level

Table 7: Personal Characteristics and Change in Social Distancing Measures

| | Did not Stay Home (=1 if yes) | Days out past week (#) | Did not Maintain Six Feet (=1 if yes) | Social Distancing Index[a] | Wears a Mask (=1 if yes) |
|---|---|---|---|---|---|
| | (1) | (2) | (3) | (4) | (5) |
| Age | -0.00 | -0.00 | -0.00 | -0.01 | 0.00 |
| | (0.00) | (0.00) | (0.00) | (0.00) | (0.00) |
| Female (=1 if yes) | -0.02 | 0.17 | 0.04 | 0.07 | -0.04 |
| | (0.05) | (0.18) | (0.06) | (0.13) | (0.05) |
| Currently Married (=1 if yes) | 0.08 | 0.40** | 0.06 | 0.28** | -0.11* |
| | (0.06) | (0.18) | (0.06) | (0.13) | (0.05) |
| Caucasian (=1 if yes) | -0.01 | 0.41** | 0.01 | 0.14 | 0.04 |
| | (0.06) | (0.21) | (0.07) | (0.15) | (0.06) |
| Schooling | -0.00 | -0.00 | 0.00 | -0.00 | -0.02 |
| | (0.02) | (0.06) | (0.02) | (0.04) | (0.02) |
| College Graduate (=1 if yes) | 0.09 | 0.32 | 0.07 | 0.29 | 0.06 |
| | (0.08) | (0.28) | (0.09) | (0.20) | (0.08) |
| Republican (=1 if yes) | 0.01 | 0.06 | 0.10 | 0.12 | 0.01 |
| | (0.07) | (0.24) | (0.08) | (0.17) | (0.07) |
| Earnings (Logged) | -0.00 | -0.10 | -0.02 | -0.05 | -0.03 |
| | (0.03) | (0.08) | (0.03) | (0.06) | (0.03) |
| Population Density (Logged) | -0.01 | -0.02 | -0.01 | -0.02 | -0.03*** |
| | (0.01) | (0.04) | (0.01) | (0.03) | (0.01) |
| Concerned About Health[b] | 0.01 | -0.01 | 0.00 | 0.01 | 0.01 |
| | (0.01) | (0.03) | (0.01) | (0.02) | (0.01) |
| COVID19 Infection Fears on a Daily Basis[c] | 0.08 | -0.26 | 0.11* | 0.10 | 0.00 |
| | (0.06) | (0.19) | (0.06) | (0.13) | (0.05) |
| Number Comorbidities | -0.04 | 0.05 | 0.04 | 0.01 | -0.02 |
| | (0.04) | (0.16) | (0.05) | (0.11) | (0.05) |
| Cognitive Score[d] | 0.03 | -0.01 | 0.01 | 0.04 | -0.04* |
| | (0.03) | (0.08) | (0.03) | (0.06) | (0.02) |
| Risk Aversion Index[e] | 0.01 | -0.01 | 0.00 | 0.01 | 0.00 |
| | (0.01) | (0.03) | (0.01) | (0.02) | (0.01) |
| Time Preference Index[f] | -0.00 | 0.00 | 0.00 | 0.00 | 0.00 |
| | (0.00) | (0.00) | (0.00) | (0.01) | (0.00) |
| Time Inconsistent (=1 if yes)[g] | 0.04 | 0.11 | 0.01 | 0.10 | 0.04 |
| | (0.05) | (0.18) | (0.06) | (0.12) | (0.05) |
| Positive Reciprocity Index | -0.01 | -0.07* | -0.01 | -0.04 | -0.00 |
| | (0.01) | (0.04) | (0.01) | (0.03) | (0.01) |
| Altruism Index | -0.00 | -0.00 | 0.01 | 0.00 | 0.01 |
| | (0.01) | (0.04) | (0.01) | (0.03) | (0.01) |

(Continued on next page)

**Table 7 (Continued):** Personal Characteristics and Change in Social Distancing Measures

|  | Did not Stay Home (=1 if yes) | Days Out Past Week (#) | Did not Maintain Six Feet (=1 if yes) | Social Distancing Index | Wears a Mask (=1 if yes) |
|---|---|---|---|---|---|
|  | (1) | (2) | (3) | (4) | (5) |
| Exponential Growth Bias Index[h] | -0.02 | -0.05 | 0.00 | -0.04 | -0.01 |
|  | (0.01) | (0.45) | (0.01) | (0.03) | (0.01) |
| Risk Misperception (in percent)[i] | -0.00 | 0.50 | 0.10 | 0.25 | 0.21 |
|  | (0.18) | (0.60) | (0.20) | (0.43) | (0.18) |
| Openness Index[j] | -0.00 | -0.12 | 0.02 | -0.02 | -0.04* |
|  | (0.02) | (0.08) | (0.02) | (0.05) | (0.02) |
| Extraversion Index[j] | -0.02 | 0.10 | 0.02 | 0.03 | 0.00 |
|  | (0.02) | (0.07) | (0.02) | (0.05) | (0.02) |
| Conscientiousness Index[j] | -0.01 | 0.13* | 0.01 | 0.04 | 0.02 |
|  | (0.02) | (0.07) | (0.02) | (0.05) | (0.02) |
| Agreeableness Index[j] | -0.01 | -0.21*** | -0.02 | -0.10* | 0.02 |
|  | (0.02) | (0.08) | (0.03) | (0.06) | (0.02) |
| Neuroticism Index[j] | 0.01 | -0.03 | -0.00 | 0.00 | -0.01 |
|  | (0.02) | (0.08) | (0.03) | (0.6) | (0.02) |
| Baseline Mean | 0.70 | 3.03 | 0.47 | 0.00 | 0.70 |
| Controls | Yes | Yes | Yes | Yes | Yes |
| R-squared | 0.01 | 0.01 | 0.02 | 0.01 | 0.01 |
| Observations | 623 | 623 | 623 | 623 | 623 |

*Notes:* Clustered standard errors are reported in parenthesis. All models include state and regional fixed effects. Behavior outcomes based on baseline wave of survey. (a) The social distancing index is based on a PCA procedure based on the four distinct proxies (did not maintain six feet in previous week; did not stay home in previous week; days out of home in previous week) of social distancing. (b) Worry about own health: 0=not at all, 10=a lot. (c) COVID-19 Fear is constructed as a binary measure based on the respondent indicating that he/she worries on daily to the following question "How frequently have you feared being infected with COVID-19 in the past month?" (d) Cognitive score is derived based on total possible correct answers (=4) divided by time allocation for task (two minutes). (e) Risk preference index from 0 (extremely risk tolerant) to 10 (extremely risk averse). (f) Time preference index from 1 (extremely impatient) to 32 (extremely patient) based on a near-frame question. Impatient (=1 if time preference index is 1). Patient (=1 if time preference index is 32). (g) "Time-inconsistent" is defined with respect to two "money" questions: 1) a near-frame question prompting individuals to choose between 160 USD today and a larger amount of USD in 12 months. 2) a distant-frame question prompting individuals to choose between 160 USD in 12 months and a larger amount of USD in 24 months. Individuals were identified as "hyperbolic" time-inconsistent if they preferred the immediate reward in the near-term frame combined with the choice of the delayed reward in the distance frame. (h) The exponential growth bias index is based on the number of incorrect responses on five exponential growth-related questions. Higher values on the index indicate higher exponential growth bias. (i) Risk misperception is the difference between the subjectively elicited case fatality rate in own county minus the actual case fatality rate at the time of the interview based on data from the Centers for Disease Control and Prevention. (j) The five variables (openness, extraversion, conscientiousness, agreeableness, neuroticism) are based on a taxonomy from psychology theory for the "Big Five" personality traits. They are indices (higher values on the index indicate increased tendency of the particular trait).
***Significant at the 1 percent level.
**Significant at the 5 percent level.
*Significant at the 10 percent level

**Table 8.** Determinants of Risk Perception of Case Fatality Rate in County of Residence

|  | Subjective Risk Perception (Percent) | Risk Misperception, Subjective-Objective (Percent) |
|---|---|---|
|  | (1) | (2) |
| Age | -0.00 | -0.00 |
|  | (0.00) | (0.00) |
| Female (=1 if yes) | 0.02* | 0.02* |
|  | (0.01) | (0.01) |
| Currently Married (=1 if yes) | -0.01 | -0.01 |
|  | (0.01) | (0.01) |
| Caucasian (=1 if yes) | -0.06*** | -0.06*** |
|  | (0.01) | (0.01) |
| Schooling | -0.00 | -0.00 |
|  | (0.00) | (0.00) |
| College Graduate (=1 if yes) | -0.00 | -0.00 |
|  | (0.01) | (0.02) |
| Republican (=1 if yes) | -0.02** | -0.03** |
|  | (0.01) | (0.01) |
| Earnings (Logged) | -0.01** | -0.01 |
|  | (0.00) | (0.00) |
| Population Density (Logged) | 0.00 | -0.00* |
|  | (0.00) | (0.00) |
| Concerned About Health | 0.00*** | 0.00** |
|  | (0.00) | (0.00) |
| COVID19 Infection Fears on a Daily Basis | -0.00 | -0.00 |
|  | (0.01) | (0.01) |
| Number Comorbidities | -0.01 | -0.00 |
|  | (0.01) | (0.01) |
| News Source Fox/MSNBC (=1 if yes) | 0.00 | 0.00 |
|  | (0.01) | (0.01) |
| COVID-19 Cases in Social Circle | -0.00 | -0.00 |
|  | (0.00) | (0.00) |
| Cognitive Score | 0.00 | 0.00 |
|  | (0.00) | (0.00) |
| Risk Aversion Index | 0.00 | 0.00 |
|  | (0.00) | (0.00) |
| Time Preference Index | -0.00* | -0.00 |
|  | (0.00) | (0.00) |
| Time-inconsistent (=1 if yes) | -0.01 | -0.01 |
|  | (0.01) | (0.01) |
| Positive Reciprocity | -0.00 | -0.00 |
|  | (0.00) | (0.00) |
| Altruism | 0.00 | 0.00 |
|  | (0.00) | (0.00) |
| Exponential Growth Bias Index | 0.01*** | 0.01*** |
|  | (0.00) | (0.00) |
| Openness Index | 0.00 | 0.00 |
|  | (0.00) | (0.00) |
| Extraversion Index | 0.00 | 0.00 |
|  | (0.00) | (0.00) |
| Conscientiousness Index | 0.01* | 0.01 |
|  | (0.00) | (0.00) |
| Agreeableness Index | -0.00 | -0.00 |
|  | (0.00) | (0.00) |
| Neuroticism Index | 0.00 | 0.00 |
|  | (0.00) | (0.00) |
| Baseline Mean | 0.08 | 0.01 |
| Controls | Yes | Yes |
| R-squared | 0.15 | 0.13 |
| Observations | 901 | 901 |

*Notes:* Clustered standard errors in parenthesis. All models include state and regional fixed effects. Full details on variable definitions are in the notes to Table VI. ***Significant at the 1 percent level. **Significant at the 5 percent level. *Significant at the 10 percent level.

**Table 9.** Determinants of Exponential Growth Bias Index

| | Exponential Growth Bias Index | | |
|---|---|---|---|
| | (1) | (2) | (3) |
| Age | 0.02*** | 0.02*** | 0.02*** |
| | (0.00) | (0.00) | (0.00) |
| Female (=1 if yes) | 0.50*** | 0.43*** | 0.32** |
| | (0.13) | (0.14) | (0.14) |
| Currently Married (=1 if yes) | -0.14 | -0.10 | -0.09 |
| | (0.14) | (0.14) | (0.14) |
| Caucasian (=1 if yes) | -0.03 | -0.01 | -0.03 |
| | (0.10) | (0.16) | (0.16) |
| Schooling | -0.07** | -0.07** | -0.06** |
| | (0.03) | (0.03) | (0.03) |
| Earnings (Logged) | -0.18*** | -0.15*** | -0.13*** |
| | (0.06) | (0.06) | (0.06) |
| Population Density (Logged) | -0.03 | -0.03 | -0.02 |
| | (0.03) | (0.03) | (0.03) |
| Concerned About Health | -0.00 | -0.01 | -0.02 |
| | (0.02) | (0.02) | (0.02) |
| COVID19 Infection Fears on a Daily Basis | 0.17 | 0.11 | 0.06 |
| | (0.14) | (0.14) | (0.14) |
| Number Comorbidities | 0.08 | 0.09 | 0.10 |
| | (0.12) | (0.11) | (0.11) |
| Cognitive Score | | -0.16*** | -0.16*** |
| | | (0.06) | (0.06) |
| Risk Aversion Index | | -0.05** | -0.06** |
| | | (0.02) | (0.02) |
| Time Preference Index | | -0.02*** | -0.02*** |
| | | (0.01) | (0.01) |
| Time-inconsistent (=1 if yes) | | -0.16 | -0.14 |
| | | (0.13) | (0.14) |
| Positive Reciprocity | | -0.04 | -0.04 |
| | | (0.03) | (0.03) |
| Altruism | | 0.05 | 0.04 |
| | | (0.03) | (0.03) |
| Risk Misperception (in percent) | | | 1.33*** |
| | | | (0.45) |
| Openness Index | | | 0.04 |
| | | | (0.06) |
| Extraversion Index | | | 0.06 |
| | | | (0.06) |
| Conscientiousness Index | | | 0.07 |
| | | | (0.05) |
| Agreeableness Index | | | 0.08 |
| | | | (0.06) |
| Neuroticism Index | | | 0.17*** |
| | | | (0.06) |
| | | | |
| Baseline Mean | 3.13e-08 | 3.13e-08 | 3.13e-08 |
| Controls | Yes | Yes | Yes |
| R-squared | 0.12 | 0.12 | 0.12 |
| | | | |
| Observations | 901 | 901 | 901 |

*Notes:* Clustered standard errors in parenthesis. All models include state and regional fixed effects. Full details on variable definitions are in the notes to Table VI. ***Significant at the 1 percent level. **Significant at the 5 percent level. *Significant at the 10 percent level.